# The Graphical Identification for Total Effects by using Surrogate Variables

**Manabu Kuroki**
Systems Innovation Dept.
Osaka University
Toyonaka, Osaka, Japan
mkuroki@sigmath.es.osaka-u.ac.jp

**Zhihong Cai**
Biostatistics Dept.
Kyoto University
Sakyo-ku, Kyoto, Japan
cai@pbh.med.kyoto-u.ac.jp

**Hiroki Motogaito**
Systems Innovation Dept.
Osaka University
Toyonaka, Osaka, Japan
h_motogt@sigmath.es.osaka-u.ac.jp

## Abstract

Consider the case where cause-effect relationships between variables can be described as a directed acyclic graph and the corresponding linear structural equation model. This paper provides graphical identifiability criteria for total effects by using surrogate variables in the case where it is difficult to observe a treatment/response variable. The results enable us to judge from graph structure whether a total effect can be identified through the observation of surrogate variables.

## 1 INTRODUCTION

### 1.1 BACKGROUND

Evaluation of causal effects from observational studies is one of the central aims in many fields of practical science. Researchers have attempted to clarify cause-effect relationships and to evaluate the causal effect of a treatment variable on a response variable through statistical data. Statistical causal analysis is a powerful tool to achieve this purpose.

The central aim of statistical causal analysis is to evaluate causal effects through both qualitative causal information and statistical data. Statistical causal analysis started with path analysis (Wright, 1923, 1934), and advanced to structural equation models (Wold, 1954; Bollen, 1989). Recently, Pearl (2000) has developed a new framework of causal modeling based on a directed acyclic graph and the corresponding nonparametric structural equation model. In addition, he provided the mathematical definition of causal effects in this framework. Furthermore, Pearl (2000) suggested several graphical identifiability conditions for causal effects, such as the back door criterion, the front door criterion and so on, which enable us to evaluate causal effects from nonexperimental data.

When a linear structural equation model is assumed as the data generating process, Tian (2004) stated that the identifiability criteria for causal effects in the nonparametric structural equation model can be translated into the identifiability criteria for total effects. Here, a "total effect" is a measure for evaluating causal effects, and can be interpreted as the change of the mean of a response variable when a treatment variable is changed by one unit through an external intervention. In addition, as a generalization of the instrumental variable method (Bowden and Turkington, 1984), Brito and Pearl (2002) developed the conditional instrumental variable method in order to identify total effects. However, it should be noted that both a treatment variable and a response variable are assumed to be observed in their discussion. Then, in the case where a treatment/response variable can not be observed, it is difficult to use their results to evaluate total effects.

In practice, there are many situations where a treatment/response variable can not be observed, and what we can measure are some surrogate variables affected by the unobserved variables. For example, in Danish longitudinal study on work, unemployment and health (Ditlevsen at al, 2005), we are interested in evaluating the effect of cynical hostility (exposure) on self-reported symptom load (response). However, both cynical hostility and symptom load are latent variables. In this study, cynical hostility was measured by 8 items derived from the Cook-Medley Hostility Scale (Everson et al, 1997), and symptom load was measured by 13 questions on physiological and mental symptoms. Then, how to use these surrogate variables to evaluate the effect of cynical hostility on symptom load is a problem to be solved. Another example is in the field of quality control. During the non-destructive testing, it may be infeasible or impracticable to observe the final characteristic unless destructing the product. However, in order to achieve quality improvement, it is necessary to understand the mechanism for how the adjustment of the treatment can influence the response (Kuroki and Miyakawa, 2004). Although it is difficult to observe the final characteristic, usually we can observe some surrogate variables that are affected by the final characteristic. Then the problem arises that whether we can evaluate the influence of treatment adjustment on the final characteristic through these surrogate variables.

The examples above show the importance of evaluating causal effects even when a treatment/response variable is unobserved, but it appears that there is a scarcity in studies of such situations. Then, the aim of this paper is to provide identifiability criteria for total effects from observational studies where a treatment/response variable can not be observed, but some surrogate variables are observed.

In this paper, we assume that cause-effect relationships between variables can be described as a directed acyclic graph and the corresponding linear structural equation model. Then, we provide graphical identifiability criteria for total effects by using surrogate variables in observational studies. We consider two cases: (1) a treatment or a response variable is not observed, and (2) both the treatment and the response variables are not observed. In each case, several surrogate variables of the unobserved variable(s) are observed. The results help us test from graph structure whether total effects can be evaluated through the observation of surrogate variables when it is difficult to observe a treatment/response variable.

### 1.2 PROBLEM DESCRIPTION

Let us consider a simple case described in Fig.1.

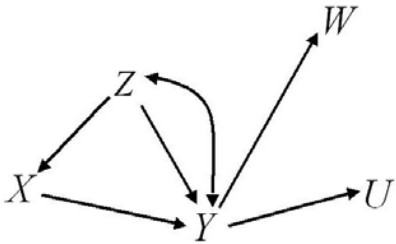

Fig.1: Path diagram (1)

Here, the target quality characteristic $Y$ can be observed only through the destructive testing, which is unreasonable in practice. When $X$ represents a treatment variable that can be controlled by an operator, suppose that we are interested in the total effect $\tau_{yx}$ of $X$ on $Y$. In order to identify the $\tau_{yx}$, some measurable characteristics, which are affected by $Y$, are utilized as surrogate variables. In Fig.1, they are denoted as $U$ and $W$. In addition, $Z$ in Fig.1 represents a set of observed covariates, such as environment factors. Furthermore, a bidirected arrow "$\longleftrightarrow$" indicates the presence of unobserved confounders affecting on two observed variables.

In this paper, such cause-effect relationships are supposed to be described as the path diagram shown in Fig.1 and the corresponding linear structural equation model. Then, we hope to determine a set of variables used in order to identify the $\tau_{yx}$ based on the graph structure shown in Fig.1. By the result of this paper, it is shown that the $\tau_{yx}^2$ is identifiable through the observation of $Z$, $X$ and a set $\{U, W\}$ of variables affected by $Y$.

## 2 PRELIMINARIES

### 2.1 PATH DIAGRAM

In statistical causal analysis, a directed acyclic graph that represents cause-effect relationships is called a path diagram. A directed graph is a pair $G = (\boldsymbol{V}, \boldsymbol{E})$, where $\boldsymbol{V}$ is a finite set of vertices and the set $\boldsymbol{E}$ of arrows is a subset of the set $\boldsymbol{V} \times \boldsymbol{V}$ of ordered pairs of distinct vertices. The graph theoretic terminology used in this paper is provided in Lauritzen (1996) and Kuroki and Cai (2004).

**DEFINITION 1 (PATH DIAGRAM)**

Suppose a directed acyclic graph $G = (\boldsymbol{V}, \boldsymbol{E})$ with a set $\boldsymbol{V} = \{V_1, \cdots, V_n\}$ of variables is given. The graph $G$ is called a path diagram, when each child-parent family in the graph $G$ represents a linear structural equation model

$$V_i = \sum_{V_j \in \mathrm{pa}(V_i)} \alpha_{v_i v_j} V_j + \epsilon_{v_i}, \qquad i = 1, \ldots, n, \quad (1)$$

where $\mathrm{pa}(V_i)$ denotes a set of parents of $V_i$ in $G$ and $\epsilon_{v_1}, \ldots, \epsilon_{v_n}$ are assumed to be independent and normally distributed. In addition, $\alpha_{v_i v_j}(\neq 0)$ is called a path coefficient. □

In a linear structural equation model discussed in this paper, it is assumed that each variable in $\boldsymbol{V}$ has mean 0 and variance 1. For further details of linear structural equation models, refer to Bollen (1989).

Here, we define some notations for future use. Let $\sigma_{xy \cdot z} = cov(X, Y | Z)$, $\sigma_{yy \cdot z} = var(Y | Z)$ and $\beta_{yx \cdot z} = \sigma_{xy \cdot z} / \sigma_{xx \cdot z}$. For disjoint sets $\boldsymbol{X}$, $\boldsymbol{Y}$ and $\boldsymbol{Z}$, let $B_{yx \cdot z}$ and $\Sigma_{xy \cdot z}$ be the regression coefficient matrix of $\boldsymbol{x}$ in the regression model of $\boldsymbol{Y}$ on $\boldsymbol{x} \cup \boldsymbol{z}$ and the conditional covariance of $\boldsymbol{X}$ and $\boldsymbol{Y}$ given $\boldsymbol{Z}$, respectively. Further, letting $\boldsymbol{X} = \{X_1, \cdots, X_q\}$, denote $\Sigma_{xx \cdot z}$ as the conditional covariance matrix of $\boldsymbol{X}$ given $\boldsymbol{Z}$, and $\sigma^{x_i x_j \cdot z}$ as the $(i, j)$ component of the inverse matrix of $\Sigma_{xx \cdot z}$.

When $\boldsymbol{Z}$ is an empty set, $\boldsymbol{Z}$ is omitted from these arguments. The similar notations are used for other parameters.

For a set $\boldsymbol{Z}$ of variables not including descendants of $V_j$, if $\boldsymbol{Z}$ d-separates $V_i$ from $V_j$ in the graph obtained by deleting from a graph $G$ an arrow pointing from $V_i$ to $V_j$, then $\beta_{v_j v_i \cdot z} = \alpha_{v_j v_i}$ holds true. This criterion is called "the single door criterion" (e.g. Pearl, 2000). In addition, when $Z$ d-separates $V_i$ from $V_j$ in the graph $G$, then $V_i$ is conditionally independent of $V_j$ given $Z$ in the corresponding distribution (e.g. Pearl, 1988, 2000).

## 2.2 LEMMAS

In this section, we introduce the back door criterion (e.g. Pearl, 2000) and the conditional instrumental variable method (Brito and Pearl, 2002) as the identifiability criteria for total effects, where a total effect $\tau_{yx}$ of $X$ on $Y$ is defined as the total sum of the products of the path coefficients on the sequence of arrows along all directed paths from $X$ to $Y$. When a total effect can be determined uniquely from the correlation parameters of observed variables, it is said to be identifiable, that is, it can be estimated consistently.

### DEFINITION 2 (BACK DOOR CRITERION)

Let $\{X, Y\}$ and $\boldsymbol{T}$ be disjoint subsets of $\boldsymbol{V}$ in a path diagram $G$. If a set $\boldsymbol{T}$ of variables satisfies the following conditions relative to an ordered pair $(X, Y)$ of variables, then $\boldsymbol{T}$ is said to satisfy the back door criterion relative to $(X, Y)$.

1. No vertex in $\boldsymbol{T}$ is a descendant of $X$, and

2. $\boldsymbol{T}$ d-separates $X$ from $Y$ in $G_{\underline{X}}$,

where $G_{\underline{X}}$ is the graph obtained by deleting from a graph $G$ all arrows emerging from vertices in $X$. □

If a set $\boldsymbol{T}$ of observed variables satisfies the back door criterion relative to $(X, Y)$ in a path diagram $G$, then the total effect $\tau_{yx}$ of $X$ on $Y$ is identifiable through the observation of $\{X, Y\} \cup \boldsymbol{T}$, and is given by the formula $\beta_{yx \cdot t}$ (Pearl, 2000).

### DEFINITION 3 (CONDITIONAL INSTRUMENTAL VARIABLE)

Let $\{X, Y, Z\}$ and $\boldsymbol{T}$ be disjoint subsets of $\boldsymbol{V}$ in a path diagram $G$. If a set $\boldsymbol{T} \cup \{Z\}$ of variables satisfies the following conditions relative to an ordered pair $(X, Y)$ of variables, then $Z$ is said to be a conditional instrumental variable given $\boldsymbol{T}$ relative to $(X, Y)$.

1. $\boldsymbol{T}$ is a subset of nondescendants of $Y$ in $G$,

2. $\boldsymbol{T}$ d-separates $Z$ from $Y$ but not from $X$ in $G_{\underline{X}}$. □

When $Z$ is a conditional instrumental variable given $\boldsymbol{T}$ relative to $(X, Y)$, a total effect of $X$ on $Y$ is identifiable through the observation of $\{X, Y, Z\} \cup \boldsymbol{T}$, and is given by $\sigma_{yz \cdot t}/\sigma_{xz \cdot t}$ (Brito and Pearl, 2002).

Regarding the discussion about selection of identifiability criteria, refer to Kuroki and Cai (2004).

In addition, in order to prove our results, the following lemmas are needed:

### LEMMA 1

When $\{X, Y\} \cup \boldsymbol{S} \cup \boldsymbol{T}$ are normally distributed, the following equations hold true:

$$\beta_{yx \cdot s} = \beta_{yx \cdot st} + B_{yt \cdot xs} B_{tx \cdot s}. \qquad (2)$$
$$\Sigma_{tt \cdot x} = \Sigma_{tt \cdot xs} + B_{ts \cdot x} \Sigma_{ss \cdot x} B'_{ts \cdot x}. \qquad (3)$$
□

Equation (2) is the result of Cochran (1938). Equation (3) is a well-known result (e.g. Whittaker, 1990).

In addition, the following lemma was given by Wermuth (1989):

### LEMMA 2

When $\{X, Y\} \cup \boldsymbol{S} \cup \boldsymbol{T}$ are normally distributed, if $\boldsymbol{T}$ is conditionally independent of $X$ given $\boldsymbol{S}$ or $Y$ is conditionally independent of $\boldsymbol{T}$ given $\{X\} \cup \boldsymbol{S}$, then $\beta_{yx \cdot st} = \beta_{yx \cdot s}$ holds true. □

## 3 IDENTIFIABILITY CRITERIA

In general, it is difficult to use the results of Pearl (2000), Brito and Pearl (2002) and Tian (2004) in order to test whether total effects can be estimated from statistical data, in the case where a treatment/response variable can not be observed. Thus, it is necessary to propose new identifiability criteria for total effects in such a situation. In order to achieve this aim, we provide the following theorem:

### THEOREM 1

Suppose that a set $\{X, U, W\} \cup \boldsymbol{Z} \cup \boldsymbol{T}$ of observed variables and an unobserved variable $Y$ satisfy the following conditions in the moral graph of $\{X, Y, U, W\} \cup \boldsymbol{Z} \cup \boldsymbol{T}$ corresponding to a path diagram $G$:

(1) $\{Y\} \cup \boldsymbol{Z} \cup \boldsymbol{T}$ separates each element $\{X, U, W\}$ from the others,

(2) Letting $\boldsymbol{R}_1$ be a subset of $\{X, U, W\}$, $\{Y, X, U, W\} \cup \boldsymbol{Z} \setminus \boldsymbol{R}_1$ separates $\boldsymbol{R}_1$ from $\boldsymbol{T}$.

(3) Letting $\boldsymbol{R}_2$ be a subset of $\{X, U, W\} \cup \boldsymbol{T}$, $\{Y, X, U, W\} \cup \boldsymbol{T} \setminus \boldsymbol{R}_2$ separates $\boldsymbol{R}_2$ from $\boldsymbol{Z}$.

Then, letting $\boldsymbol{S} = \{X, U, W\} \cup \boldsymbol{Z} \cup \boldsymbol{T}$, $\Sigma_{sy}\Sigma_{ys}$ is identifiable. □

Here, elements of the $\Sigma_{ss}^{-1}$, $\sigma^{uw}$, $\sigma^{xw}$ and $\sigma^{xu}$ are assumed to be non-zeros in Theorem 1. In addition, condition (2) is omitted from Theorem 1 when $\boldsymbol{T}$ is an empty set, and condition (3) is omitted from Theorem 1 when $\boldsymbol{Z}$ is an empty set.

Theorem 1 is based on a necessary and sufficient condition for a single factor models, which is provided in Vicard (2000).

## PROOF OF THEOREM 1

Letting $\boldsymbol{S} = \{X, U, W\} \cup \boldsymbol{Z} \cup \boldsymbol{T}$, from Lemma 1,

$$\Sigma_{ss} = \Sigma_{ss \cdot y} + \frac{1}{\sigma_{yy}} \Sigma_{sy} \Sigma_{ys},$$

Here, from the Sherman-Morrison-Woodbury formula for matrix inversion (e.g. Rao, 1973), we can obtain

$$\Sigma_{ss}^{-1} = \Sigma_{ss \cdot y}^{-1} - \frac{\Sigma_{ss \cdot y}^{-1} \Sigma_{sy} \Sigma_{ys} \Sigma_{ss \cdot y}^{-1}}{\sigma_{yy} + \Sigma_{ys} \Sigma_{ss \cdot y}^{-1} \Sigma_{sy}} = \Sigma_{ss \cdot y}^{-1} - \boldsymbol{\lambda} \boldsymbol{\lambda}', \quad (4)$$

where $\boldsymbol{\lambda} = \Sigma_{ss \cdot y}^{-1} \Sigma_{sy} / \sqrt{\sigma_{yy} + \Sigma_{ys} \Sigma_{ss \cdot y}^{-1} \Sigma_{sy}}$. When partitioning $\boldsymbol{\lambda}'$ into $(\lambda_1, \lambda_2, \lambda_3, \boldsymbol{\lambda}_4', \boldsymbol{\lambda}_5')$ according to elements of $\boldsymbol{S}$, from condition (1), by using Lemmas 1 and 2,

$$\Sigma_{ss}^{-1} = \Sigma_{ss \cdot y}^{-1} - \boldsymbol{\lambda} \boldsymbol{\lambda}'$$

$$= \begin{pmatrix} \sigma^{xx \cdot y} & 0 & 0 & \Sigma^{xz \cdot y} & \Sigma^{xt \cdot y} \\ & \sigma^{uu \cdot y} & 0 & \Sigma^{uz \cdot y} & \Sigma^{ut \cdot y} \\ & & \sigma^{ww \cdot y} & \Sigma^{wz \cdot y} & \Sigma^{wt \cdot y} \\ & & & \Sigma^{zz \cdot y} & \Sigma^{zt \cdot y} \\ & & & & \Sigma^{tt \cdot y} \end{pmatrix}$$

$$- \begin{pmatrix} \lambda_1^2 & \lambda_1 \lambda_2 & \lambda_1 \lambda_3 & \lambda_1 \boldsymbol{\lambda}_4' & \lambda_1 \boldsymbol{\lambda}_5' \\ & \lambda_2^2 & \lambda_2 \lambda_3 & \lambda_2 \boldsymbol{\lambda}_4' & \lambda_2 \boldsymbol{\lambda}_5' \\ & & \lambda_3^2 & \lambda_3 \boldsymbol{\lambda}_4' & \lambda_3 \boldsymbol{\lambda}_5' \\ & & & \boldsymbol{\lambda}_4 \boldsymbol{\lambda}_4' & \boldsymbol{\lambda}_4 \boldsymbol{\lambda}_5' \\ & & & & \boldsymbol{\lambda}_5 \boldsymbol{\lambda}_5' \end{pmatrix}, \quad (5)$$

where the lower triangular components are omitted. From equation (5), since $\lambda_1 \lambda_2 = -\sigma^{xu}$, $\lambda_1 \lambda_3 = -\sigma^{xw}$ and $\lambda_2 \lambda_3 = -\sigma^{uw}$ hold true, we can obtain

$$\lambda_1^2 = -\frac{\sigma^{xu} \sigma^{xw}}{\sigma^{uw}}, \quad \lambda_2^2 = -\frac{\sigma^{xu} \sigma^{uw}}{\sigma^{xw}}$$

$$\lambda_3^2 = -\frac{\sigma^{xw} \sigma^{uw}}{\sigma^{xu}}.$$

Thus, from condition (2), we can obtain

if $\Sigma^{tx \cdot y} = \boldsymbol{0}$, then $\boldsymbol{\lambda}_5 = -\frac{1}{\lambda_1} \Sigma^{tx}$,

if $\Sigma^{tu \cdot y} = \boldsymbol{0}$, then $\boldsymbol{\lambda}_5 = -\frac{1}{\lambda_2} \Sigma^{tu}$,

if $\Sigma^{tw \cdot y} = \boldsymbol{0}$, then $\boldsymbol{\lambda}_5 = -\frac{1}{\lambda_3} \Sigma^{tw}$.

In addition, from condition (3), we can obtain if $\Sigma^{zx \cdot y} = \boldsymbol{0}$, then $\boldsymbol{\lambda}_4 = -\frac{1}{\lambda_1} \Sigma^{zx}$, if $\Sigma^{zu \cdot y} = \boldsymbol{0}$, then $\boldsymbol{\lambda}_4 = -\frac{1}{\lambda_2} \Sigma^{zu}$, if $\Sigma^{zw \cdot y} = \boldsymbol{0}$, then $\boldsymbol{\lambda}_4 = -\frac{1}{\lambda_3} \Sigma^{zw}$, if $\Sigma^{zt \cdot y} = \boldsymbol{0}$, then $\boldsymbol{\lambda}_4 = -\frac{\Sigma^{zt} \boldsymbol{\lambda}_5}{\boldsymbol{\lambda}_5' \boldsymbol{\lambda}_5}$.

By noting $\lambda_1/\lambda_2 = \sigma^{xw}/\sigma^{uw}$, $\lambda_1/\lambda_3 = \sigma^{xu}/\sigma^{uw}$ and $\lambda_2/\lambda_3 = \sigma^{xu}/\sigma^{xw}$, $\boldsymbol{\lambda}\boldsymbol{\lambda}'$ is identifiable. Thus, from equation (4), $\Sigma_{ss \cdot y}$ is also identifiable and is given by $\Sigma_{ss \cdot y} = (\Sigma_{ss}^{-1} + \boldsymbol{\lambda} \boldsymbol{\lambda}')^{-1}$. In addition, letting $\sigma_{yy} = 1$, we can obtain $\Sigma_{sy} \Sigma_{ys} = \Sigma_{ss} - \Sigma_{ss \cdot y}$, which indicates that $\Sigma_{sy} \Sigma_{ys}$ is identifiable. Q.E.D.

Here, consider the path diagram shown in Fig.2.

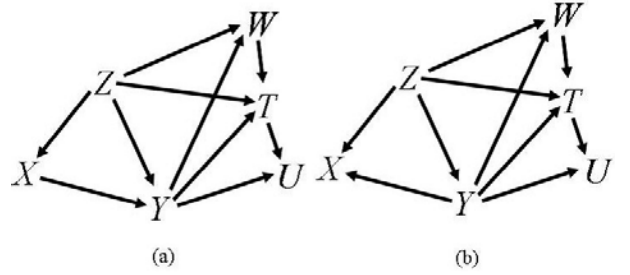

Fig. 2: Path Diagram (2)

Fig.2 (a) represents the case where the response variable $Y$ is unobserved, but surrogate variables $W$, $U$ and $\boldsymbol{T}$ affected by $Y$ are observed. In addition, $Z$ satisfies the back door criterion relative to $(X, Y)$. First, we can obtain

$$\sigma_{xy \cdot z}^2 = (\sigma_{xy} - \Sigma_{xz} \Sigma_{zz}^{-1} \Sigma_{zy})^2$$
$$= \sigma_{xy}^2 - 2\sigma_{xy} \Sigma_{yz} \Sigma_{zz}^{-1} \Sigma_{xz}$$
$$+ \Sigma_{xz} \Sigma_{zz}^{-1} \Sigma_{zy} \Sigma_{yz} \Sigma_{zz}^{-1} \Sigma_{zx},$$

which is identifiable since both $\Sigma_{zy} \Sigma_{yz}$ and $\sigma_{xy}^2$ are identifiable from Theorem 1. Thus, when $Z$ satisfies the back door criterion relative to $(X, Y)$, since the total effect $\tau_{yx}$ of $X$ on $Y$ can be given by $\tau_{yx} = \beta_{yx \cdot z}$ and $\sigma_{xx \cdot z}$ is estimable, we can obtain

$$\tau_{yx}^2 = \beta_{yx \cdot z}^2 = \frac{\sigma_{xy \cdot z}^2}{\sigma_{xx \cdot z}^2},$$

which indicates that the square of the total effect $\tau_{yx}^2$ is identifiable.

Fig.2 (b) represents the case where the treatment variable $Y$ is unobserved, but surrogate variables $W$, $U$ and $\boldsymbol{T}$ affected by $Y$ are observed. In addition, $Z$ satisfies the back door criterion relative to $(X, Y)$, Then,

$$\sigma_{yy \cdot z} = \sigma_{yy} - \Sigma_{yz} \Sigma_{zz}^{-1} \Sigma_{zy}$$
$$= \sigma_{yy} - \text{tr}(\Sigma_{zz}^{-1} \Sigma_{zy} \Sigma_{yz}),$$

which is also identifiable from Theorem 1. Thus, when $Z$ satisfies the back door criterion relative to $(Y, X)$, since the total effect $\tau_{xy}$ of $Y$ on $X$ can be given by $\tau_{xy} = \beta_{xy \cdot z}$, we can obtain

$$\tau_{xy}^2 = \beta_{xy \cdot z}^2 = \frac{\sigma_{xy \cdot z}^2}{\sigma_{yy \cdot z}^2},$$

which indicates that the square of the total effect $\tau_{xy}^2$ is identifiable.

Next, consider the path diagram shown in Fig.3.

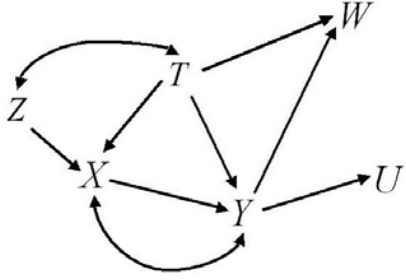

Fig. 3: Path Diagram (3)

Fig.3 represents the case where the response variable $Y$ is unobserved, but surrogate variables $W$ and $U$ affected by $Y$ are observed. Thus, when $Z$ is the conditional instrumental variable given $T$ relative to $(X, Y)$, since the total effect $\tau_{yx}$ of $X$ on $Y$ can be given by $\tau_{yx} = \sigma_{yz \cdot t}/\sigma_{xz \cdot t}$, we can obtain

$$\begin{aligned}\sigma_{yz \cdot t}^2 &= (\sigma_{yz} - \Sigma_{xt}\Sigma_{tt}^{-1}\Sigma_{ty})^2 \\ &= \sigma_{yz}^2 - 2\sigma_{yz}\Sigma_{yt}\Sigma_{tt}^{-1}\Sigma_{tx} \\ &\quad + \Sigma_{xt}\Sigma_{tt}^{-1}\Sigma_{ty}\Sigma_{yt}\Sigma_{tt}^{-1}\Sigma_{tx},\end{aligned}$$

which is identifiable from Theorem 1. Thus, since $\sigma_{xz \cdot t}$ is estimable, we can understand that $\tau_{yx}^2$ is identifiable.

In addition, we consider the case where both a treatment variable and a response variable are unobserved.

**THEOREM 2**

Suppose that a set $\{U_1, U_2, W_1, W_2\} \cup \boldsymbol{T}_1 \cup \boldsymbol{T}_2 \cup \boldsymbol{Z}$ of observed variables and an unobserved variable $\{X_1, X_2\}$ satisfy the following conditions:

(1) Letting $X = U_1$, $Y = X_1$, $W = W_1$, $S = U_2$ and $\boldsymbol{T} = \boldsymbol{T}_1$, $\{X, Y, S, W\} \cup \boldsymbol{T} \cup \boldsymbol{Z}$ satisfies conditions (1), (2), and (3) in Theorem 1,

(2) Letting $X = U_1$, $Y = X_2$, $S = U_2$, $W = W_2$ and $\boldsymbol{T} = \{W_1\} \cup \boldsymbol{T}_2$, $\{X, Y, S, W\} \cup \boldsymbol{T} \cup \boldsymbol{Z}$ satisfies conditions (1), (2) and (3) in Theorem 1, and

(3) $\{X_1\} \cup \boldsymbol{Z}$ d-separates $U_1$ from $X_2$.

Then, $\beta_{x_2 x_1 \cdot z}^2$ is identifiable. □

Here, $\boldsymbol{T}_1 \cap \boldsymbol{T}_2 = \phi$ is not required in Theorem 2.

Theorem 2 can be applied to the path diagram shown in Fig.4 with unobserved treatment variable $X_1$ and response variable $X_2$, where $U_1$ and $W_1$ are observed surrogates of $X_1$, and $U_2$ and $W_2$ are observed surrogates of $X_2$. Here, for simplicity we assume $\boldsymbol{T}_i$ ($i = 1, 2$) is an empty set in this graph.

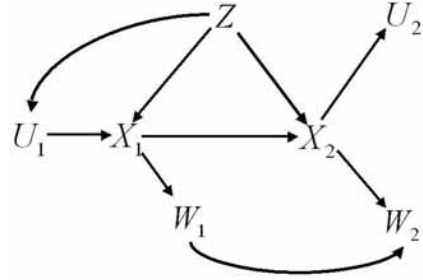

Fig. 4: Path Diagram (4)

**PROOF OF THEOREM 2**

From condition (1), letting $\boldsymbol{S}_1 = \{U_1, W_1, U_2\} \cup \boldsymbol{T}_1 \cup \boldsymbol{Z}$, $\Sigma_{s_1 x_1} \Sigma_{x_1 s_1}$ is identifiable from Theorem 1. In addition, from condition (2), letting $\boldsymbol{S}_2 = \{U_1, U_2, W_1, W_2\} \cup \boldsymbol{T} \cup \boldsymbol{Z}$, $\Sigma_{s_2 x_2} \Sigma_{x_2 s_2}$ is identifiable from Theorem 1. Since $\beta_{x_2 u_1 \cdot z} = \beta_{x_2 x_1 \cdot z} \beta_{x_1 u_1 \cdot z}$ holds true from condition (3), $\beta_{x_2 x_1 \cdot z}^2$ is identifiable from Thorem 1.
Q.E.D.

As an example, we test whether the total effect of $X_1$ on $X_2$ is identifiable through the observation of $\{U_1, U_2, W_1, W_2, Z\}$ in the path diagram shown in Fig.4. First, letting $X = U_1$, $Y = X_1$, $U = U_1$, $W = W_1$ and $\boldsymbol{T} = \phi$, $\{X_1, W_1, U_1, U_2, Z\}$ satisfies conditions (1), (2) and (3) in Theorem 1. Next, letting $X = U_1$, $Y = X_2$, $U = U_2$, $W = W_2$ and $\boldsymbol{T} = \{W_1\}$, $\{X_2, W_1, W_2, U_1, U_2, Z\}$ satisfies conditions (1), (2) and (3) in Theorem 1. Furthermore, $\{X_1, Z\}$ d-separates $U_1$ from $X_2$. Thus, the square of the total effect of $X_1$ on $X_2$ is identifiable from Theorem 2. Here, it is noted that condition (3) is not required in the case where $U_1$ is the conditional instrumental variable given $Z$ relative to $(X_1, X_2)$.

## 4 CONCLUSION

This paper derived the graphical identifiability criteria for total effects in observational studies with an unobserved treatment/response variable, when cause-effect relationships between variables can be described as a directed acyclic graph and the corresponding linear structural equation model. The results of this paper enable us to evaluate total effects when it is difficult to observe a treatment/response variable but graphical structure of cause-effect relationships between variables is known.

Finally, we would like to provide some comments on our results. First, regarding the case where a treatment variable or response variable is not observed, the graphical identifiability criteria for total effects proposed in this paper are related to that of a single factor model proposed by Stanghellini (1997) and Vicard (2000) and the path model with one hidden variable by Stanghellini and Wermuth (2005). However, it should be noted that we are interested in the identification of total effects but not in that of all the path coefficients, while Stanghellini and Wermuth (2005) are interested in all the path coefficients with only one latent variable. Second, regarding the case where neither a treatment variable nor a response variable is observed, it is possible to extend Theorem 2 to the case where there are several unobserved variables, which is closely connected to the identification of multi-factor models with correlated errors. Thus, as the future works, the detailed discussion regarding the identification of multi-factor models (e.g. Grzebyk et al., 2004) is required.

**ACKNOWLEGDEMENT**

Thanks go to Kazushi Maruo of Osaka University for his helpful comments on this paper. The comments of the reviewers on preliminary versions of this paper are also acknowledged. This research was supported by the Sumitomo Foundation, the Murata Overseas Scholarship Foundation, the College Women's Association of Japan and the Ministry of Education, Culture, Sports, Science and Technology of Japan.